\documentclass[aps,prl,twocolumn]{revtex4}
\usepackage{times}
\usepackage{latexsym}
\usepackage{graphicx}
\usepackage{verbatim,times,bbm}
\usepackage{color}
\usepackage{appendix}

\def\uno{\mbox{1 \kern-.59em {\rm l}}}

\def\be{\begin{equation}}
\def\ee{\end{equation}}
\def\ba{\begin{eqnarray}}
\def\ea{\end{eqnarray}}
\def\lo{\longrightarrow}

\def\la{\langle}
\def\ra{\rangle}
\def\a{\alpha}
\def\b{\beta}

\begin{document}

\title{Systematics of Entanglement Distribution by Separable States}

\author{V. Karimipour, L. Memarzadeh and Najmeh Tabe Bordbar}
\affiliation{Department of Physics, Sharif University of Technology, Teheran, Iran}

\begin{abstract}
We show that three  distant labs  A, B and C, having no prior entanglement can establish a shared GHZ state, by mediating two particles which remain separable from each other and from all the other parties throughout the process.  The success probability is $\frac{1}{7}$. We prove this in a general framework for systematic distribution of entangled states between 2 and more parties with separable states in $d$ dimensions.  The proposed method may facilitate the construction of multi-node quantum networks and many other processes which use multi-partite entangled states. 
\end{abstract}

\pacs{03.67.Bg,  03.67.Mn
}

\maketitle

{\bf Introduction\ } Decades of study of entanglement, from its inception in relation with conceptual framework of the quantum theory \cite{EPR1, EPR2, EPR3}, to its recent upsurge as a tool in quantum computation and quantum information science \cite{q.computation} and the inevitable need for its quantification has shown that its various unexplored aspects are still intriguing and can still surprise us. In the history of study of entanglement, the turning point was the discovery that entanglement can be used to teleport quantum states \cite{tel},  to share secret keys and to do dense coding, all pointing to the fact that this concept is a useful resource, as important as many other physical quantities.  Since then it has been shown  that entanglement can be manipulated \cite{manipulation}, measured \cite{measurement},  distributed\cite{distribution} and distilled \cite{distillation}. The distinctive feature of entanglement is that it cannot be created by local action and classical communication (LOCC) \cite{Bennett}.  Any attempt for distributing entanglement  between two or more distant particles, should involve either a direct  interaction between the particles when they are close \cite{parametricdownconversion} or should be mediated between them by sending a mediating particle through a quantum channel \cite{MP}. This latter method is of course very much vulnerable to noise, since the state of mediating particle is very fragile. This problem becomes very severe if we note that the mediating particle may be entangled with the distant particles.\\
 
It came as a big surprise when it was shown \cite{EDSS} that a mediating particle (C)  can be entirely separable from the distant particles (A and B) which are to be entangled.  It was then shown that such a protocol is also possible with  Gaussian states \cite{mista1, mista2} where the actual experiments with these continuous variable states were reported in \cite{peutinger, vollmer} and for qubit states of light in \cite{fedrizzi}.  The question of what type of initial qubit states can lead to this phenomena was explored in \cite{AKay} and the cost for this entanglement increase was related to the discord \cite{vedral, zurek} in the initial state in \cite{Bruss, pater}.  \\

However important questions have remained unanswered until now: Can we distribute multi-partite entanglement between $N$ parties by using separable ancillas? If yes, how? How do we distribute $d-$ level maximally entangled states by using separable ancillas? Can we provide a systematic method for these protocols? These questions are not only of conceptual relevance, specially in view of the fact that $GHZ$ states \cite{ghz} have a different kind of non-locality than Bell states, but they have practical relevance in view of their role in many quantum protocols like quantum secret sharing \cite{BHB, kari} and other quantum protocols using quantum networks \cite{leweinstein}. Moreover in view of the studies in \cite{Bruss} and \cite{pater} on the cost of entanglement sharing and its relation with quantum correlations, a systematic generalization of this phenomena to $d$ levels and $N$ parties, will certainly lead to deeper insight on these inter-relations and the bounds that they suggest. In this letter we provide positive answers to all these questions.  We first discuss $d-$ level Bell state distribution and then go on to GHZ distribution between many parties.  \\   
 
{\bf {Notations and conventions:} } 
\begin{itemize} 
\item{} In all the multi-partite states, we write the order of parties only in the left hand side of the states which is meant to be the same as in the right hand side. For example we write the same state as $|\Psi\ra_{ABC}=|011\ra$ or $|\Psi\ra_{BCA}=|110\ra $, since by necessity we have to change the order of terms in the states, the reader is asked to always check the order of party labels on the left hand side.
A CNOT operation on $C$, controlled by $A$ is denoted by $C_{_{AC}}$ and acts as $C_{_{AC}}|i,j\ra_{_{AC}}=|i,i+j\ra$. The combined states before Alice CNOT operation(s), after her CNOT operation(s) and after Bob CNOT operations are denoted respectively by  $\rho^{(0)}$, $\rho^{(1)}$ and $\rho^{(2)}$. \\
\end{itemize}

{\bf Distribution of Bell states in arbitrary dimensions\ }
Consider two parties, Alice(A) and Bob(B), who possess pure states of the form 
\be
|\phi(k)\ra_A=\frac{1}{\sqrt{d}}\sum_{j=0}^{d-1}\omega^{s_jk}|j\ra, \ \ \ \  |\phi(-k)\ra_B=\frac{1}{\sqrt{d}}\sum_{j=0}^{d-1}\omega^{-s_jk}|j\ra.
\ee
where $K$ is an integer, $\omega=e^{\frac{2\pi i}{K}}$ is a phase,  $k=0,1,\cdots K-1$,  and the non-negative integers  $s_j$ are all different from each other and satisfy the following conditions: \\

i) $0\leq s_i < \frac{K}{2}$\\

ii) for any four indices $i,j,m$ and $n$, which are not all equal 
\be s_i+s_j= s_m+s_n,\  \  \  \  {\rm only \ \ if } \    \  \  (i,j)=(m,n) \  \   {\rm or} \  \  \ (i,j)=(n,m).\ee
These conditions will play an important role in the process as we will see. In particular we note the consequences

\ba\nonumber\label{condit}
&&\frac{1}{K}\sum_{k=0}^{K-1}\omega^{(s_i-s_j)k}=\delta_{i,j},\cr
&&\frac{1}{K}\sum_{k=0}^{K-1}\omega^{(s_i+s_j-s_m-s_n)k}=\delta_{i,m}\delta_{j,n}+\delta_{i,n}\delta_{j,m}-\delta_{ij}\delta_{mn}\delta_{jm}.
\ea

Alice has also a $d$-level ancilla  $C$ at state $|0\ra_C$. She now enacts the operation $C_{_{AC}}$,  turning the fully separable state $|\phi(k)\ra_{_A} |0\ra_{_C} |\phi(-k)\ra_{_B} $ into the following partially entangled state, where $C$ is entangled to $A$ but not to $B$: 
  
\be\label{d-dim-phi}
|\Phi^{(1)}(k)\ra_{_{ACB}}=\frac{1}{d}\sum_{j=0,l=0}^{d-1} \omega^{(s_j-s_l)k}|j,j,l\ra,
\ee
 
and sends the ancilla $C$ to B where he performs $C^{-1}_{_{BC}}$, turning the state to 
\ba\label{d-dim-psi}
|\Phi^{(2)}(k)\ra_{_{CAB}}&=&\frac{1}{d}\sum_{j=0,l=0}^{d-1} \omega^{(s_j-s_l)k}|j-l,j,l\ra\cr &=&\frac{1}{\sqrt{d}}\sum_{m=0}^{d-1} |m\ra|\chi_{_m}(k)\ra
\ea
where
\be
|\chi_{_m}(k)\ra_{_{AB}}=\frac{1}{\sqrt{d}}\sum_{j=0}^{d-1}\omega^{(s_j-s_{j-m})k}|j,j-m\ra,
\ee
are maximally entangled states. The important point, as we will see, is that $|\chi_{_0}\ra$ and only this state, is independent of $k$. \\

If now the ancilla  qudit $C$ is measured in the computational basis, the state of $AB$ is projected onto the maximally entangled state $|\chi_{_0}\ra_{AB}$ with a given probability. In this way A and B, have used the ancilla $C$ to entangle their initially separable state $|\phi(k)\ra_A|\phi(-k)\ra_B$ into a maximally entangled one.  The problem however is that the ancilla has been entangled to the states during the process. We now show that this entanglement can entirely be removed and the qudit C can be completely separated from the state of the two parties $AB$ in all stages of the process.\\

To do this we use the fact that a mixture of entangled states can be separable and use two consecutive processes, namely mixing and symmetrization.  The two main stages of the process correspond to the two states $|\Phi^{(1)}(k)\ra_{_{ACB}}$ ( after Alice CNOT operation) and $|\Phi^{(2)}(k)\ra_{_{CAB}}$ (after Bob inverse CNOT operation). If we disentangle $C$ in each of  these two states, it means that $C$ has been a separable ancilla in its travel between Alice and Bob during  the whole process.\\

Consider first the states $|\Phi^{(1)}(k)\ra$ in (\ref{d-dim-phi}) and determine the  un-normalized mixture, 
\be
\rho'^{(1)}_{_{ACB}}=\sum_{k=0}^{K-1} |\Phi^{(1)}(k)\ra\la \Phi^{(1)}(k)|
\ee
which by using the conditions on $\{s_i\} 's $ becomes 
\be\label{}
\rho'^{(1)}_{_{ACB}}= \frac{K}{d^2}\left(\sum_{j\ne l} \pi_{jjl} + d |GHZ\ra\la GHZ|\right)
\ee
where $\pi_{ijk}=|ijk\ra\la ijk|$ and 
\be
|GHZ\ra_{_{ACB}}=\frac{1}{\sqrt{d}}\sum_{j}|jjj\ra.
\ee
In view of the GHZ term the ancilla $C$ has not been made separable from AB.  However we now note that by 
 the very operation of Alice on C, which involves only A and C,  this state has the structure of $\rho_{_{AC|B}}$, that is $B$ is certainly separable from AC. Therefore if we make it symmetric with respect to the interchange $B\leftrightarrow C$, then it will have also the structure $AB|C$ and the ancilla C will be separable from the original particles A and B. The GHZ state has obviously this symmetry, so we add the projectors $\pi_{jlj}$ to this state and obtain the separable state  
 \be
 \rho'^{(1)}_{_{ACB}}\lo \rho'^{(1)}_{_{ACB}}+\frac{K}{d^2}\sum_{j\ne l}\pi_{jlj}
 \ee
which after proper normalization gives us the separable density matrix after operation of Alice and before Bob operation
\be\label{rho1}
\rho^{(1)}_{_{ACB}}=\frac{1}{2d-1}|GHZ\ra\la GHZ|+\frac{1}{d(2d-1)}\sum_{j\ne l}(\pi_{jjl}+\pi_{jlj}).
\ee
From this state and by noting that it can be written with both forms of indices $ACB$ or $ABC$ (due to its symmetry) and by inverting the CNOT operation of Alice, we will obtain the fully separable original state of Alice, Bob and the ancilla:
\ba
\rho^{(0)}_{_{A|B|C}}&=&\frac{d}{K(2d-1)}\sum_{k=0}^{K-1}|\phi(k)\ra\la \phi(k)|\otimes |\phi(-k)\ra\la \phi(-k)|\otimes |0\ra\la 0| \cr &+&\frac{1}{d(2d-1)}\sum_{j\ne l}\pi_{j,j,l-j}
\ea
From (\ref{rho1}) we now obtain the final state after Bob's inverse CNOT operation, 
\ba
\rho^{(2)}_{_{AB|C}}&=&\frac{1}{2d-1} |\chi_0\ra\la \chi_0|\otimes |0\ra\la 0| \cr &+& \frac{1}{d(2d-1)}\sum_{j\ne l}(\pi_{jl,j-l}+\pi_{jj,l-j}).
\ea
This shows that when the ancilla is measured in the computational basis, the state of AB is projected onto a Bell state with probability $\frac{1}{2d-1}$. This probability is independent of the number of states $K$. This number is determined once we have chosen the parameters $s_j$ satisfying the two conditions. We illustrate this with a few examples:\\

{\it Example 1- qubits}:  The simplest assignment is to take $s_0=0$ and $s_1=1$. With $K=3$, we obtain $\omega=e^{\frac{2\pi i}{3}}$, where Alice and Bob each use  three states. 
Another solution is given if we take $K=4$ or  $\omega=e^{2\frac{\pi i}{4}}$, where Alice and Bob each use 4 states. This leads to the state given in \cite{EDSS}. \\

{\it Example 2- qutrits}: The simplest assignment is $s_0=0, s_1=1, s_2=3$. The smallest number of states is $K=7$ with 
 $\omega=e^{\frac{2\pi i}{7}}$. \\

{\it Example 3- four-level states}: The  simplest assignment is $s_0=0, s_1=1, s_2=3, s_3= 7$.  The smallest number of states is $K=15$. \\

{\it Example 4- $d$ level states}: The simplest assignment is given by $s_i=2^i-1$ and $K=2^d-1$ states. \\

\begin{figure}
\begin{center}
\vskip -3 mm
\includegraphics[width=8cm, height=6cm]{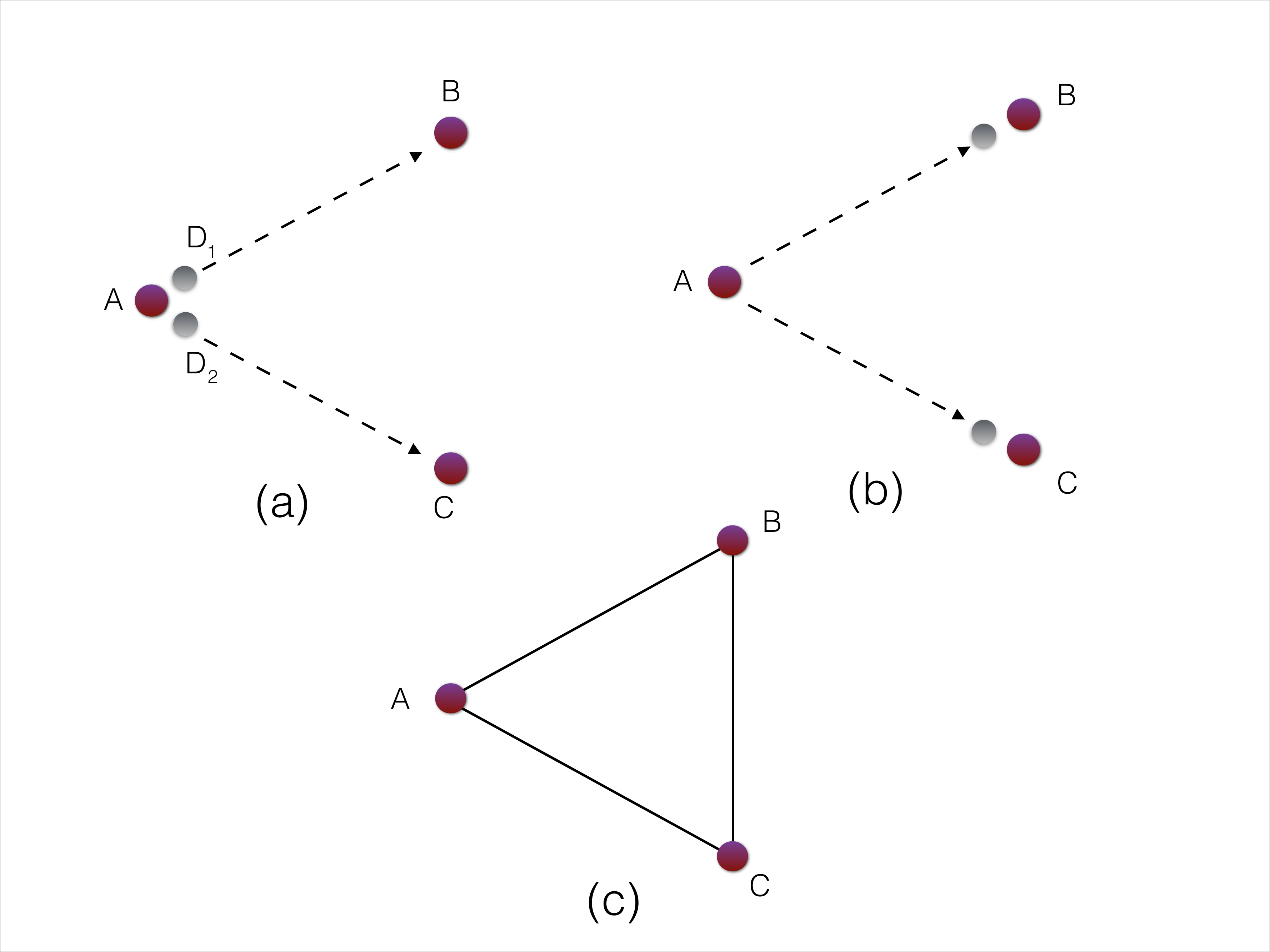}
\caption{(Color online) GHZ distribution with separable states. (a) Alice operates on two ancillas with her CNOT operations, controlled by her qubit A and sends the two ancillas to Bob and Charlie. (b) they operate on their respective ancillas by their CNOT's (inverse CNOT's in case of qudits).  (c) At the end a GHZ state is shared between A, B and C with a probability $P_{success}=\frac{1}{7}.$ Throughout the process the ancillas remain separable from each other and from ABC. The process can be generalized for $N$ parties and for qudits. }
\label{GHZ3fig}
\end{center}
\end{figure}

{\bf Distribution of GHZ states\ }
We now use the same strategy to distribute GHZ state between three parties, figure (\ref{GHZ3fig}). The idea is again mixing and symmetrization of product states. To simplify the presentation we explain in detail the idea for 2-level GHZ states between 3 parties. This simple setting allows us to understand where the conditions on the state parameters come from and how one should proceed for other cases.  In view of the previous section, generalization to $d$ level states and more parties is straightforward, although its presentation will clutter the text with many indices.  

‎‎‎‎
Here Alice, Bob and Charlie possess pure states of the form 
$\frac{1}{\sqrt{2}}(|0\ra+\a_k|1\ra)$, $\frac{1}{\sqrt{2}}(|0\ra+\b_k|1\ra),$ and $\frac{1}{\sqrt{2}}( |0\ra+\gamma_k|1\ra)$ respectively, where $\a_k\b_k\gamma_k=1$. Alice has two ancillas $D_1$ and $D_2$ which are both at the states $|0\ra$. Alice now performs the operations $C_{_{AD_1}}$ and $C_{_{AD_2}}$ and turn the product state into 
\ba\label{GHZ-phi}
&&|\Phi^{(1)}(k)\ra_{_{AD_1D_2,BC}}=\cr &&\frac{1}{\sqrt{8}}(|000\ra+\a_k|111\ra)(|0\ra+\b_k|1\ra)(|0\ra+\gamma_k|1\ra). 
\ea
and then sends the ancilla qubits $D_1$ and $D_2$ to Bob and Charlie who perform their respective CNOT operations (inverse CNOT for  $d-$ level states) and turn the state into

\ba\label{GHZ-psi}
|\Phi^{(2)}_k\ra_{_{D_1D_2,ABC}}&=&\frac{1}{2}(|00\ra|\chi_{_{00}}(k)\ra+|01\ra|\chi_{_{01}}(k)\ra\cr &+&|10\ra|\chi_{_{10}}(k)\ra+|11\ra|\chi_{_{11}}(k)\ra)
\ea

where 
\ba\nonumber
|\chi_{_{00}}\ra&=&\frac{1}{\sqrt{2}}(|000\ra+|111\ra), \cr 
|\chi_{_{01}}\ra&=&\frac{1}{\sqrt{2}}(\gamma|001\ra+\gamma^{-1} |110\ra),\cr
|\chi_{_{10}}\ra&=&\frac{1}{\sqrt{2}}(\beta|010\ra+\b^{-1}|101\ra), \cr  |\chi_{_{11}}\ra&=&\frac{1}{\sqrt{2}}(\a|100\ra+\a^{-1}|011\ra). 
\ea
Note that $|\chi_{00}\ra$ is the standard $GHZ_3$ state and it is important that it is independent of $k$.\\

Now Bob and Charlie can measure the ancillas $D_1$ and $D_2$ in the computational basis which projects their joint state with A into the GHZ state $|\chi_{_{00}}\ra$ with a given probability (to be determined later). To remove the entanglement of the two ancillas $D_1$ and $D_2$ from the rest of states we again use the procedure of mixing and symmetrization. We first mix the final state (\ref{GHZ-psi})  and form  $\rho'^{(2)}_{_{D_1D_2,ABC}}:=\sum_k |\Phi^{(2)}(k)\ra\la\Phi^{(2)}(k) |$. To make this state separable we require that  
\be
\sum_{k} |\chi_{ij}(k)\ra\la \chi_{i'j'}(k)|=0\ \ \ \ \ \ \  (i,j)\ne (i',j'),
\ee
which leads to the independent equations:
\be\label{cons-GHZ-psi}
\sum_{k}\a_k=\sum_{k}\b_k=\sum_{k}\gamma_k=\sum_{k}\frac{\a_k}{\b_k}=\sum_{k}\frac{\a_k}{\gamma_k}=\sum_{k}\frac{\b_k}{\gamma_k}=0.
\ee
Under these conditions, the mixed state will be 

\be\rho'^{(2)}_{_{D_1D_2\mid ABC}}=\frac{1}{4}|00\ra\la 00|\otimes |GHZ_3\ra\la GHZ_3|.\ee
We will later come back to this state. 
The next step is to mix the states (\ref{GHZ-phi}) after Alice's and before Bob's operation.   So we form the un-normalized mixture  $\rho'^{(1)}_{_{AD_1D_2,BC}}:=\sum_k|\Phi^{(1)}(k)\ra\la \Phi^{(1)}(k)|$ and symmetrize it with respect 
 to the interchange $(D_1D_2\leftrightarrow BC)$. Since by the very operation of Alice on the ancillas alone, this state has the structure $\rho_{_{AD_1D_2\mid BC}}$, if we can make it symmetric, it will have also the structure of $\rho_{_{ABC\mid D_1D_2}}$, hence the ancilla will be separable from $ABC$.  Using an abbreviated notation as ${\cal D}\equiv D_1D_2$ and ${\cal B}\equiv BC$ and also using the binary notations for two-qubit states (e.g. $|\overline{3}\ra\equiv |11\ra$) we can write 
\ba\label{new-GHZ-phi}
|\Phi^{(1)}(k)\ra_{_{A{\cal D}{\cal B}}}&=&\frac{1}{\sqrt{8}}(|0\overline{0}\overline{0}\ra+\gamma|0\overline{0}\overline{1}\ra+\beta|0\overline{0}\overline{2}\ra+\beta\gamma|0\overline{0}\overline{3}\ra\cr
+\a |1\overline{3}\overline{0}\ra&+&\a\gamma|1\overline{3}\overline{1}\ra+\a\beta|1\overline{3}\overline{2}\ra+|1\overline{3}\overline{3}\ra).
\ea
In view of this new form of the state $|\Phi^{(1)}(k)\ra$, we note that the non-symmetric terms in $\rho'^{(1)}$ are of two kinds: 1) those which are projectors like $\pi_{0\overline{0}\overline{2}}$ and 2) those which are cross terms like $|0\overline{0}\overline{0}\ra\la 0\overline{0}\overline{2}|$. We demand that all these cross terms vanish. A glimpse at (\ref{new-GHZ-phi}) shows that this requires that 
\be\label{cons-GHZ-phi}
\sum_k \a_k^2=\sum_k \b_k^2=\sum_k \gamma_k^2=0,
\ee
where we have also used condition (\ref{cons-GHZ-psi}). 
We are then left with 
$$
\rho'^{(1)}_{_{A{\cal D}{\cal B}}}=\frac{K}{8}\left[2|GHZ_5\ra\la GHZ_5| + \sum_{i=1}^3\pi_{0\overline{0}\overline{i}}+\sum_{i=0}^2\pi_{1\overline{3}\overline{i}})\right]
$$
which can be symmetrized by adding suitable projectors. After normalization the state will be 
\ba\nonumber
\rho^{(1)}_{_{A{\cal B}{\cal D}}}&=&\frac{1}{7}|GHZ_5\ra\la GHZ_5| \cr &+& \frac{1}{14}\left[\sum_{i=1}^3(\pi_{0\overline{0}\overline{i}}+\pi_{0\overline{i}\overline{0}})+\sum_{i=0}^2(\pi_{1\overline{3}\overline{i}}+\pi_{1\overline{i}\overline{3}})\right].
\ea
Note from this state that the two particles $D_1$ and $D_2$, are not only separable from A, B and C, but are also separable from each other. \\

We should now consider what effect this addition of projectors has on the initial $\rho^{(0)}$ and $\rho^{(2)}$.  In view of the fact that the effect of $CNOT$'s by Alice and Bob turn any projector into another projector, we see that by this addition the initial state $\rho^{(0)}$ of Alice, Bob and Charlie and also their final states $\rho^{(2)}$ remain separable and the probability of distributing a $GHZ_3$ states is given by  $P_{success}=\frac{1}{7}$. \\

It remains to solve the constraints (\ref{cons-GHZ-psi}) and (\ref{cons-GHZ-phi}). To do this we use the relation $\a_k\b_k\gamma_k=1$ to express everything in terms of $\a_k$ and $\beta_k$ and make the ansatz that $\b_k=\a_k^2$ to transform the constraints (\ref{cons-GHZ-psi}) and (\ref{cons-GHZ-phi}) into 

\be
\sum_k\a_k^{r}=0,\ \ \ \ \ \  r=1,2,\cdots 6.
\ee This is simply solved by taking $\a_k=e^{\frac{2\pi i k}{7}}$ which then gives 
\be
\a_k=e^{\frac{2\pi i k}{7}},\ \ \ \  \b_k=e^{\frac{4\pi i k}{7}},\ \ \ \  \gamma_k=e^{\frac{8\pi i k}{7}},
\ee
and $k$ ranges from 0 to $6$, that is, Alice, Bob and Charlie each should use 7 different states of the form given in the beginning of this section. \\

The same analysis can be carried out for  $N$ parties. In fact for $N$ parties, each party $A_j$, $ \ \ j=1,\ 2, \ \cdots N$,  should have $2^N-1$ states of the form  
\be
|\phi_{j}(k)\ra_{A_j}=\frac{1}{\sqrt{2}}(|0\ra+\omega^{jk}|1\ra), \ \ \ \ \  k=0, 1, 2, \cdots 2^N-2,
\ee
where $\omega = e^{\frac{2\pi i}{2^N-1}}$.
It is easy to show that running this protocol in the same way as before leads to sharing a GHZ state between the parties with probability $P_{success}=\frac{1}{2^N-1}$.

In summary, we have introduced a constructive method for using separable states to distribute d-level Bell states between 2 parties and GHZ states between 3 and more parties. The method is straightforward to generalize to $GHZ_N$ sharing for qudits. Several questions remain for future investigations: First one can consider other classes of multi-partite states, like W states, 1-uniform states \cite{zyk} and  cluster states \cite{clus} which are important for one-way quantum computation. Second one may see how GHZ state sharing with separable states can be implemented with Gaussian variables along the the theoretical lines set in \cite{mista1, mista2} and their experimental realization in \cite{peutinger, vollmer, fedrizzi}. Finally,  investigation of the cost of multi-partite entanglement sharing \cite{Bruss}, \cite{pater} and its relation with multi-partite  quantum correlation  present in the initial states, will surf again, a  study which may lead to a host of  connections and bounds between two not-well understood concepts. \\

We thank Shakib Vedaei for interesting discussions. 




\end{document}